\documentclass[doublecol]{epl2}

\usepackage{graphicx}
\usepackage{amssymb}
\usepackage{dcolumn}
\usepackage{bm}
\usepackage{epstopdf}
\usepackage{float,epsfig}

\title{Effects of spin-flip scattering on gapped Dirac fermions}

\author{S. A. Yang\inst{1} \and Z. Qiao\inst{1} \and Y. Yao\inst{2}
\and J. Shi\inst{2} \and Q. Niu\inst{1}}
\shortauthor{S. A. Yang \etal}

\institute{
  \inst{1} Department of Physics, The University of Texas, Austin,
Texas, 78712-0264, USA\\
  \inst{2} Institute of Physics, Chinese Academy of Sciences,
Beijing 100190, China
}


\abstract{
We investigate the effects of spin-flip scattering on the Hall
transport and the spectral properties of gapped Dirac fermions. We find
that in the weak scattering regime, the Berry curvature distribution
is dramatically compressed in the electronic energy spectrum,
becoming singular at band edges. As a result the Hall conductivity
has a sudden jump (or drop) of $e^2/(2h)$ when the Fermi energy
sweeps across the band edges, and otherwise is a constant quantized
in units of $e^2/(2h)$. The spectral properties such as the
density of states and the spin polarization are also greatly enhanced near
band edges. Possible experimental methods to detect these effects are discussed.}

\begin{document}

\maketitle

\section{Introduction}
Dirac fermion has found its ubiquitous
appearance in interesting physical systems such as
graphene\cite{graphene} and topological
insulators\cite{TI,Hazan2010}. The most important feature of Dirac
fermion is its strong (pseudo)spin-orbit coupling, which is directly
encoded in the Hamiltonian and underlies many unusual effects. When
certain symmetry breaking mechanism is introduced, a gap can be
opened up at the Dirac point.
Gapped Dirac fermion (GDF) is in some sense more interesting in
that its electronic band develops a Berry curvature gauge
field\cite{BerryBook,Xiao2010} which acts like a magnetic field in the momentum
space, and deflects electron flow in transverse directions leading
to the anomalous Hall effect\cite{Nagaosa2010}.

Disorder scattering usually has strong effects on the Hall transport
even in the weak disorder limit when the system is in a metallic
state. So far, most studies of scattering effects on Dirac fermions
have focused on spin independent scattering. However, due to the strong
spin-orbit coupling, it is natural to expect that the carrier motion
should have a sensitive dependence on the change of spin state
during a scattering.

In this paper, we study the effects of spin-flip scattering on the
properties of GDFs. Our work is motivated by recent advances in
creating GDF on the surface of topological
insulators\cite{Chen2010,Wray2010} and the ability of systematic control
of surface carrier density through doping or
gating\cite{Xia2009}. We show that in the weak scattering
regime spin-flip scattering compresses the Berry curvature
distribution dramatically towards the band edge. Hence whenever the
Fermi energy is tuned across a band edge, the Hall conductivity has
a sudden change of $e^2/(2h)$, i.e. half of the conductance quantum.
Away from the band edges, the Hall conductivity is a constant
independent of the carrier density. Since Hall transport is central
to several exotic physical effects\cite{Garate2010} proposed
recently for topological insulators, the discovery presented here is
expected to have important observable consequences.
We also find that the spectral properties like the density of states (DOS)
and the spin polarization which are usually not very sensitive to
disorder scattering get greatly enhanced near band edges by the
spin-flip scattering, which can be detected by photoemission
spectroscopy or tunneling spectroscopy measurements.

\section{Quantized Hall conductivity and Berry curvature compression}
A GDF is described by the Hamiltonian
\begin{equation}
\hat{\mathcal{H}}_0=v_F\bm{k}\cdot\bm{\tau}+\Delta\tau_z,
\end{equation}
where $v_F$ is the Fermi velocity, $\bm{k}=(k_x,k_y,0)$, and
$\bm{\tau}=(\tau_x,\tau_y,\tau_z)$ is a vector of Pauli matrices
acting on the spin degrees of freedom. Without the second term, the
Hamiltonian is for a massless Dirac fermion with a cone-like energy
dispersion. The term $\Delta\tau_z$ induces a finite mass and opens
up an energy gap of $2\Delta$ in the spectrum. Such a term can
be generated by breaking the time reversal symmetry at the surface
of a topological insulator, for example, through magnetic
doping\cite{Chen2010} or coating with insulating ferromagnetic films.

Because of the strong spin-orbit coupling and the spin splitting
induced by the mass term, an anomalous Hall effect
in the absence of external magnetic field is expected. In the weak scattering
regime, the anomalous Hall
conductivity has an important contribution $\sigma_{xy}^0$
from the momentum space Berry curvature of the spin-orbit coupled bands\cite{Nagaosa2010}.
It is known as the intrinsic contribution because it is an intrinsic
property of a crystal. For a
two-dimensional (2D) system the intrinsic contribution is given by the integral of momentum
space Berry curvature $\Omega_{n\bm{k}}$ of all the occupied states
$|n\bm{k}\rangle$ ($n$ is the band index),
$\sigma_{xy}^0=-\frac{e^2}{\hbar}\frac{1}{A}\sum_{n\bm{k}}\Omega_{n\bm{k}}f_{n\bm{k}}$,
where $A$ is the area of the 2D system, $f_{n\bm{k}}$ is the Fermi
distribution function,
\begin{equation}\Omega_{n\bm{k}}= -\sum_{n'\neq
n}\frac{2\mathrm
{Im}\langle n\bm{k}|v_x|n'\bm{k}\rangle\langle
n'\bm{k}|v_y|n\bm{k}\rangle}{(\omega_{n'\bm{k}}-\omega_{n\bm{k}})^2},
\end{equation}
$v_i$ ($=v_F\tau_i/\hbar$ with $i=x,y$ for GDF) is the velocity operator and $\hbar\omega_{n\bm{k}}$ is the
energy of the state $|n\bm{k}\rangle$. The distribution of Berry
curvature in the energy spectrum can be described by the
density of Berry curvature $\Omega(\varepsilon)\equiv\frac{1}{A}
\sum_{n\bm{k}}\Omega_{n\bm{k}}\delta(\varepsilon-\hbar\omega_{n\bm{k}})$.
Then the intrinsic contribution can be put into the form
$\sigma_{xy}^0=-\frac{e^2}{\hbar}\int d\varepsilon
\Omega(\varepsilon)f(\varepsilon)$.

\begin{figure}
\centering \epsfig{file=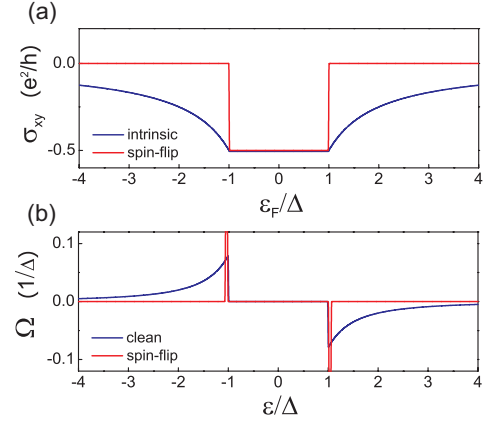,width=0.7\linewidth}
\caption{\label{fig:one} (color online). (a) The anomalous Hall
conductivity and (b) the density of Berry curvature for the gapped Dirac
fermion (with $\Delta>0$) as a function of energy. The blue curve
in (a) is for the intrinsic contribution and the red curve is for a
disordered system with weak spin-flip scattering. The Berry curvature is
compressed by the scattering to be singular at band edges, as shown
schematically in (b).}
\end{figure}

For GDFs, the density of Berry curvature and the intrinsic contribution of Hall
conductivity can be easily calculated using the formulae above. The result is
\begin{equation}
\Omega(\varepsilon)=-\frac{\Delta}{4\pi\varepsilon^2}\mathrm{sgn}(\varepsilon)\Theta(|\varepsilon|-|\Delta|),
\end{equation}
\begin{equation}
\sigma_{xy}^0(\varepsilon_F)=-\frac{e^2}{2h}[\frac{\Delta}{|\varepsilon_F|}\Theta(|\varepsilon_F|-|\Delta|)
+\mathrm{sgn}(\Delta)\Theta(|\Delta|-|\varepsilon_F|)],
\end{equation}
where $\Theta(x)$ is the Heaviside step function and $\varepsilon_F$ is
the Fermi energy. As shown by the blue curves in fig.~\ref{fig:one},
while the density of Berry curvature has its maximum (or minimum)
value at band edges, its distribution spreads over the whole energy
spectrum, hence the value of $\sigma_{xy}^0$ changes gradually as a
function of the Fermi energy. For each band the total weight of
Berry curvature corresponds to a contribution of $\pm e^2/(2h)$ to the
Hall conductivity, where the sign difference reflects the different
helicities of the two bands. It needs to be mentioned that the half quantized contribution
from the filled lower band does not contradict the well known fact that the contribution from
a completely filled band must be integer quantized\cite{Nagaosa2010}.
This is because the Dirac model is energetically unbounded. It
can only serve as a low energy effective model for any real physical systems, where
the evaluation of the contribution from all the completely filled bands obviously
has to go beyond the effective model. On the other hand, it is crucial to notice that the disorder related
contribution, which is our focus in this paper, is tied to the Fermi surface hence is well
captured within such effective model.

The existence of disorder scattering is essential for stabilizing
the electron distribution under a drive field and establishing
the dynamic steady state. In the case of the anomalous Hall effect, the role of
disorder scattering is especially important yet complicated.
In the weak scattering regime, when expanded
in terms of the disorder density $n_\mathrm{dis}$, the leading
contribution to $\sigma_{xy}$ is generally of order
$n_\mathrm{dis}^{-1}$, known as the skew scattering contribution which originates
from the asymmetry in the scattering rate and depends on the
details of the disorder model\cite{Nagaosa2010}. The
next leading order (of $n_\mathrm{dis}^{0}$ dependence) consists of the intrinsic contribution discussed
above and the side jump contribution\cite{Nagaosa2010}.
The skew scattering contribution and the side jump contribution arise from disorder
scattering, they are referred to as extrinsic contributions.
The side jump contribution is peculiar in that it arises from scattering but
is independent of both the scattering strength and the disorder density. Due to its presence,
even for disorder models with vanishing skew scattering (such as Gaussian disorder
models), the Hall conductivity $\sigma_{xy}$ in the weak disorder limit $n_\mathrm{dis}
\rightarrow 0^+$ does not correspond to the value of the intrinsic contribution $\sigma_{xy}^0$\cite{Nagaosa2010}.
This is an important point that needs to be emphasized.

In a recent study of the anomalous Hall effect\cite{Yang2011}, we find that the
extrinsic contributions to $\sigma_{xy}$ depend sensitively on the
spin structure of disorder scattering. This is understandable
because the Hall transport is a result of spin-orbit coupling, different
actions on the spin state during a scattering will also strongly
affect the carrier's orbital motion. Here we are most interested in the effects of spin-flip scattering
on the gapped Dirac fermions. Such kind of scattering can be modeled as a
random potential $U(\bm{r})=\sum_a
\bm{B}^a(\bm{r}-\bm{R}_a)\cdot\bm{\tau}=B_x(\bm{r})\tau_x+B_y(\bm{r})\tau_y$,
with $\bm{B}^a$ being a random in-plane vector. We assume that the
random field satisfies the correlation $\langle
B_i(\bm{q})B_j^*(\bm{q}')\rangle_c=n_\mathrm{dis}u^2\delta_{ij}\delta(\bm{q}-\bm{q}')$,
where $i,j\in\{x,y\}$, $B_i(\bm{q})$ is the Fourier component of
$B_i(\bm{r})$, $\langle\cdots\rangle_c$ stands for disorder average
and $u$ is the scattering strength. We assume that the system has no
preferred in-plane direction, such that the random vector $\bm{B}^a$
is uniformly distributed in plane. Then the third order disorder correlation
$\langle UUU\rangle_c$ must vanish identically. Because the skew scattering
contribution depends on the third order correlation\cite{Nagaosa2010,Sinitsyn2007},
this means that the skew
scattering process is forbidden for this kind of disorder. Therefore the
leading order contribution to the Hall conductivity comes from the
intrinsic and the side jump terms which are independent of scattering
strength and disorder density.

In the weak scattering regime, the Hall conductivity can be evaluated perturbatively
using the Kubo-Streda formula\cite{Streda1982}. The calculation has been detailed in our previous
work\cite{Yang2011}. Here we quote the final result which is that
the side jump contribution is
\begin{equation}\sigma_{xy}^\mathrm{sj}(\varepsilon_F)=\frac{e^2}{2h}\frac{\Delta}{|\varepsilon_F|}
\Theta(|\varepsilon_F|-|\Delta|),
\end{equation}
hence the total Hall conductivity is given by
\begin{equation}\label{sig}
\sigma_{xy}(\varepsilon_F)\simeq\sigma_{xy}^0+\sigma_{xy}^\mathrm{sj}=-\frac{e^2}{2h}\mathrm{sgn}(\Delta)\Theta
(|\Delta|-|\varepsilon_F|).
\end{equation}
This is a very interesting result. As plotted in fig.~\ref{fig:one}(a), $\sigma_{xy}$ has a sudden jump
or drop of half the conductance quantum at the band edges. As soon
as the Fermi energy passes the band edge, the value of $\sigma_{xy}$
becomes a constant quantized in units of $e^2/2h$. The effect of the spin-flip scattering
makes the anomalous Hall conductivity quantized in the whole energy spectrum, which is different
from other types of disorders\cite{Sinitsyn2007,Yang2011}.

In this paper, we would like to promote another perspective
on the unusual phenomena described by eq.~(\ref{sig}).
Instead of starting from a clean system and treating disorders as perturbations, we
start directly from a disordered system and try to generalize the concept of
Berry curvature for the disordered case. For each
eigenstate $|\alpha\rangle$ with eigen-energy $\hbar\omega_\alpha$ of the disordered system,
we define its Berry curvature as
\begin{equation}\label{gb}\Omega_\alpha\equiv
-\sum_{\beta\neq
\alpha}\frac{2\mathrm{Im}\langle\alpha|v_x|\beta\rangle
\langle\beta|v_y|\alpha\rangle}{(\omega_\beta-\omega_\alpha)^2}.
\end{equation}
We further introduce a Berry curvature operator
$\hat{\Omega}\equiv\sum_\alpha\Omega_\alpha|\alpha\rangle\langle\alpha|$,
such that the density of Berry curvature can be
generalized as
\begin{equation}\label{dob}
\Omega(\varepsilon)\equiv\frac{1}{A}\mathrm{Tr}
\left[\hat{\Omega}\,\delta(\varepsilon-\hat{\mathcal{H}})\right].
\end{equation}
It follows from the Kubo formula that the Hall conductivity for
disordered systems can be expressed as
\begin{equation}\label{kubo}
\sigma_{xy}=-\frac{e^2}{\hbar}\int d\varepsilon
\langle\Omega(\varepsilon)\rangle_c f(\varepsilon),
\end{equation}
with a disorder averaged density of Berry curvature. This is an exact result.
We again emphasize that although the definition
of generalized Berry curvature recovers that for a perfect crystal when the
eigenstates $|\alpha\rangle$ are simply Bloch states
$|n\bm{k}\rangle$, the weak disorder limit ($n_\mathrm{dis}\rightarrow 0^+$)
of eq.~(\ref{kubo}) is not equal to the intrinsic contribution. It also contains
the side jump contribution (for vanishing skew scattering).
Mathematically, this is
because the disorder averaged Berry curvature contains nontrivial vertex corrections which
do not vanish in the weak disorder limit.
From eq.~(\ref{sig}) we find that in
the weak disorder limit,
\begin{equation}
\langle\Omega(\varepsilon)\rangle_c=-\frac{1}{4\pi}\mathrm{sgn}(\varepsilon\Delta)\delta(|\varepsilon|-|\Delta|).
\end{equation}
As shown schematically in fig.~\ref{fig:one}(b), the density of Berry
curvature, which is originally spread over the energy spectrum, gets
compressed dramatically towards the band edges by the scattering,
while its total weight ($\pm\frac{1}{4\pi}$ for lower
and upper band respectively) is kept unchanged.

\begin{figure}
\centering \epsfig{file=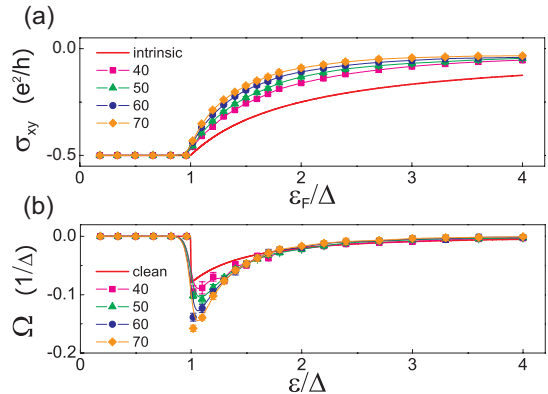,width=0.8\linewidth}
\caption{\label{fig:two} (color online). Numerical result of (a) the
Hall conductivity and (b) the Berry curvature distribution for
increasing system sizes with $N_k=40,50,60,70$ averaged over
$200,100,80,50$ samples respectively. In the calculation we set
$v_F=\Delta=1$, the momentum cutoff $K_c=6$, and the strength $|B_i(\bm{q})|$ ($i=x,y$) is a
random number uniformly distributed in an interval $[0,W]$ with
$W=0.03$ (accordingly $n_\mathrm{dis}u^2\simeq W^2K_c^2/3\sim 0.01$).
The error bar stands for one standard deviation.}
\end{figure}

Based on eqs.~(\ref{gb})-(\ref{kubo})
we perform a numerical calculation to further confirm our analytic result. Due to the particle-hole symmetry,
$\sigma_{xy}$ is a symmetric function with respect to
$\varepsilon_F=0$, hence here we only focus on the positive energy range. The calculation is performed in
momentum space on disordered systems of size $N_k\times N_k$.
Figure~\ref{fig:two} shows the results for $\sigma_{xy}$
and $\Omega$ with different system size $N_k$.
The physically meaningful value of a transport coefficient
such as $\sigma_{xy}$ corresponds to the thermodynamic limit
when $N_k\rightarrow\infty$. The result in fig.~\ref{fig:two} indeed shows
that the density of Berry curvature is
squeezed by the weak spin-flip scattering to the upper band bottom. Consequently
the contribution to the Hall conductivity from the upper band
approaches a half-quantized plateau in the energy spectrum, which confirms our analytic
result.

To completely understand the microscopic mechanism of this peculiar disorder effect is difficult.
Nevertheless, we notice that for the gapped Dirac fermion,
the spin-flip scattering has weak scattering rates near
the band bottom because states there have large spin $z$-components, meanwhile large
scattering rates occur for high energies where the spins are
more or less lying in the plane. Our result indicates that the Berry curvature tends to be expelled from
the strong scattering region towards the weak scattering region while keeping
its total weight unchanged, hence
making the curvature distribution squeezed towards
band bottom. This is similar to the migration of
Chern number carrying states driven by impurity scattering
observed in the quantum Hall effect\cite{Liu1996,Sheng2001}.

\section{Enhanced DOS and spin polarization}
\begin{figure}
\centering \epsfig{file=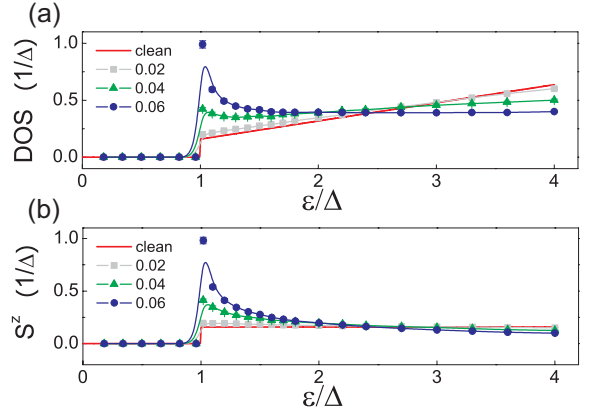,width=0.85\linewidth}
\caption{\label{fig:three} (color online). Numerical results of (a)
the DOS and (b) the spin polarization $S^z$ for increasing disorder strength
with $W=0.02,0.04,0.06$ with each data point averaged over 50 samples,
where $W$ is defined in the caption of fig.~\ref{fig:two}. In the
calculation, we set $N_k=70$, and $v_F=\Delta=1$.}
\end{figure}

Spectral properties such
as the DOS are usually considered to be insensitive to weak disorder
scattering. This is particularly true when the DOS does not vary
much in the energy range under consideration, such as for a 2D free
electron gas where the DOS is a constant. A GDF gas without disorder has a
DOS
\begin{equation}\rho_0(\varepsilon)=\frac{|\varepsilon|}{2\pi
v_F^2}\Theta(|\varepsilon|-|\Delta|).
\end{equation}
There are three salient features of this DOS that are crucial for
the effects to be
discussed below: (1) $\rho_0$ increases linearly from the band edge;
(2) $\rho_0$ has a finite value $\frac{|\Delta|}{2\pi v_F^2}$ at
band edges; (3) $\rho_0$ does not depend on the value of the spin
splitting $\Delta$ as long as $|\varepsilon|>|\Delta|$.
Figure \ref{fig:three}(a) shows the numerical result of the DOS for a
clean system together with the DOS for a disordered system with spin-flip disorders.
It is observed that spin-flip scattering strongly enhances the DOS
near the band edge, and slightly decreases the DOS at large energies,
as being required by the conservation of the total number of states.

This effect can be qualitatively understood from the self-energy correction due to scattering.
In the Born approximation, (assuming $\varepsilon>\Delta>0$ in the following),
the $2\times 2$ retarded self energy is given by
$\Sigma^R=\frac{1}{A}\sum_{\bm{k}}\langle U
G_0^R(\bm{k},\varepsilon)U\rangle_c$, with
\begin{equation}\label{selfR}
\mathrm{Re}\Sigma^R(\varepsilon)=-\frac{n_\mathrm{dis}u^2}{2\pi
v_F^2}\ln\left|\frac{\varepsilon^2-\varepsilon_c^2}{\varepsilon^2
-\Delta^2}\right|(\varepsilon\tau_0-\Delta\tau_z),
\end{equation}
and $\mathrm{Im}\Sigma^R(\varepsilon)=
-\frac{n_\mathrm{dis}u^2}{2v_F^2}(\varepsilon\tau_0-\Delta\tau_z)$,
where $\varepsilon_c$ is the energy cutoff and $\tau_0$ is the
$2\times 2$ identity matrix. While the imaginary part of the self-energy
only contributes to a small level broadening in the weak scattering
regime, the real part has a large magnitude near the band edge and
diverges logarithmically as
$\varepsilon\rightarrow\Delta$. In fact, this disorder induced
singularity is a generic feature for systems having a finite
(unperturbed) DOS at the band edge, as is the case for the GDF system.

Thermodynamic properties can be extracted from the Green function
$G^R=(\varepsilon-\mathcal{H}_0-\Sigma^R)^{-1}$. From
eq.~(\ref{selfR}), the self-energy only corrects the energy argument
$\varepsilon$ and spin splitting $\Delta$ in $G^R$. It follows that
the disordered DOS can  be approximated as
\begin{equation}\rho(\varepsilon)\simeq\rho_0[\varepsilon(1+\lambda),\Delta(1+\lambda)],
\end{equation}
with the right hand side standing for the DOS at energy
$\varepsilon(1+\lambda)$ for a clean GDF system with $\Delta$
replaced by $\Delta(1+\lambda)$, where
$\lambda=\frac{n_\mathrm{dis}u^2}{2\pi
v_F^2}\ln\left|\frac{\varepsilon^2-\varepsilon_c^2}{\varepsilon^2-\Delta^2}\right|$
is the prefactor appearing in eq.~(\ref{selfR}). Near the band edge,
$\lambda$ is a positive number, hence
$\rho(\varepsilon)\simeq(1+\lambda)\rho_0(\varepsilon)$ is enhanced
by a factor of $(1+\lambda)$ compared with the clean case\cite{DOS_cross}. Now it is
clear that a strong enhancement of DOS by scattering requires the
original DOS has a large variation on the energy scale of
$\lambda\Delta$. Therefore such effect does not occur for a
conventional 2D free electron gas, and although the GDF near the band edge resembles
the 2D free electron gas, the self-energy correction makes the Dirac behavior at higher
energies relevant. Similarly, the spin polarization
$S^z(\varepsilon)=-\frac{1}{\pi A}\mathrm{Im
Tr}[\tau_zG^R(\varepsilon)]$ is also enhanced at band edge by the factor
$(1+\lambda)$, as shown in fig.~\ref{fig:three}(b).

\section{Discussion}
The exotic effects reported here arise from the
interplay between the GDF and the spin-flip scattering. There are two
important properties of GDFs underlying these effects: the
Berry curvature and the unusual DOS. Disorder scattering
modifies these fundamental properties in the spectrum, thereby
strongly influences the system behavior.

Spin conserving scattering such as $B_0(\bm{r})\tau_0$ or
$B_z(\bm{r})\tau_z$ also affects the anomalous Hall transport of GDF
system in metallic state. In contrast with the spin-flip scattering, their skew scattering
contribution is in general non-vanishing.
Their side jump contributions are also distinct from that of the spin-flip
scattering\cite{Sinitsyn2007,Yang2011}. In particular, the
Berry curvature is not compressed into a small region in the
spectrum. The combined effects of different kinds of
scattering would depend on the competition between them\cite{Yang2011,Nomura2011}.
As for the disorder modified DOS, the spin conserving
scattering has similar effect as the spin-flip scattering. This can be
understood by noticing that the self-energy correction for spin conserving
scattering only differs from eq.~(\ref{selfR}) by a sign change of $\Delta$. Because $\rho_0$
does not depend on $\Delta$ (for energies within the bands), the result
is similar to that for the spin-flip scattering. Meanwhile, the spin
polarization becomes $S^z(\varepsilon)\simeq
(1-\lambda)S^z_0(\varepsilon)$. In energy range where $\lambda>1$,
the spin polarization can even have a sign change.

GDF has been realized on the surface of topological
insulators\cite{Chen2010}. This is partly motivated by the possibility to achieve a half-quantized anomalous
Hall effect by tuning the Fermi level into the
surface band gap\cite{Hazan2010,Chu2010}. In contrast,
our focus here is instead on the energy range outside the gap.
We notice that for a purely 2D system with
embedded gapped Dirac spectrum, it is possible to realize
half-quantized anomalous Hall effect in the metallic state with
the help of spin-flip scattering. For example, consider a Hamiltonian
$\mathcal{H}_0=v_F(\sin k_x\tau_x+\sin k_y \tau_y)+(\Delta+2-\cos
k_x-\cos k_y)\tau_z$ $(\Delta>0)$ which corresponds to a tight-binding lattice model
proposed to model the Mn doped HgTe/CdTe
quantum wells\cite{Liu2008}. It has a direct gap at $\Gamma$
point of the Brillouin zone. The low energy effective
model near the gap is just that for a GDF. As we mentioned before, the contribution to the Hall
conductivity from the completely filled lower band (or any remote valence band) must be integer
quantized. Therefore once the Fermi energy is slightly above the upper band
bottom and spin-flip scattering is turned on, an $e^2/(2h)$
contribution arises from the upper band bottom making the total Hall
conductivity half-quantized. In fact, the
non-integer value of Hall conductivity dictates that the
system must be conducting\cite{Niu1985}. The effect itself would be very
interesting since it realizes a fractional quantized Hall effect for non-interacting electrons in a
metallic state, plus it does \emph{not} require the completely
filled valence bands to have a nontrivial topology, \emph{i.e.} with a nonzero
Chern number\cite{Xiao2010}.

Finally, we briefly comment on the possible experimental methods to
detect the effects studied here on the surface of topological
insulators. The surface band gap can be opened by depositing an
ultrathin insulating ferromagnetic film on top of a topological
insulator such that a proximity-induced exchange coupling can be
introduced. Spin-flip scattering can be realized by thermally
excited spin waves in the ferromagnetic film. However, to make spin
wave scattering dominate over other scattering processes is a
nontrivial task. It requires a temperature range where spin wave
excitation dominates over phonon excitation and improved sample
quality such that very few impurities are present near the interface.
Ferromagetic insulating materials such as EuS and EuO with the Curie
temperature much lower than the Debye temperature are possible
candidates. The Hall conductivity can be measured through standard
multi-terminal setup. The sudden change of Hall conductivity at band
edges will also manifest in the signal of magneto-electric or
magneto-optic effects\cite{Garate2010}. Enhanced DOS and spin
polarization at band edges can be directly detected through
the photoemission spectroscopy or the tunneling spectroscopy measurements,
which should be much easier.

\acknowledgments
The authors thank N. A. Sinitsyn and J.
Sinova for helpful discussions. This work is supported by NSFC
(10974231) (Y.Y.), DOE (DE-FG03-02ER45958) and Texas Advanced
Research Program (Q.N.).


\begin{thebibliography}{0}
\bibitem{DOS_cross} A more careful evaluation leads to the result that
$\rho(\varepsilon)=\rho_0(\varepsilon)\left[1+
\frac{n_\mathrm{dis}u^2}{2\pi
v_F^2}\left(\ln\left|\frac{\varepsilon^2-\varepsilon_c^2}{\varepsilon^2-\Delta^2}\right|-2\right)\right]$,
which shows that the curves of $\rho(\varepsilon)$ and $\rho_0(\varepsilon)$ have a
crossing point at the energy $\varepsilon_*$ when
$\ln\left|\frac{\varepsilon_*^2-\varepsilon_c^2}{\varepsilon_*^2-\Delta^2}\right|-2=0$.
With the parameter values in our numerical calculation, we find that $\varepsilon_*=2.55$
which agrees very well with the result in Fig.\ref{fig:three}(a).
\end{thebibliography}
\end{document}